# Opportunities for multiscale computational modelling of serotonergic drug effects in Alzheimer's disease

Alok Joshi[1,*], Da-Hui Wang[2,3], Steven Watterson[4], Paula L. McClean[4], Chandan K. Behera[1], Trevor Sharp[5], and KongFatt Wong-Lin[1,*]

[1]Intelligent Systems Research Centre, Ulster University, Derry~Londonderry, Northern Ireland, UK

[2]State Key Laboratory of Cognitive Neuroscience and Learning, Beijing Normal University, Beijing, China

[3]School of System Science, Beijing Normal University, Beijing, China

[4]Northern Ireland Centre for Stratified Medicine, Biomedical Sciences Research Institute, Ulster University, Derry~Londonderry, Northern Ireland, UK

[5]Department of Pharmacology, University of Oxford, Oxford, UK

**\*Corresponding Authors**

Alok Joshi

    Address: Intelligent Systems Research Centre, Ulster University, Northland Road, Derry~Londonderry, BT48 7JL, Northern Ireland, UK

    E-mail: a.joshi@ulster.ac.uk

KongFatt Wong-Lin

    Address: Intelligent Systems Research Centre, Ulster University, Northland Road, Derry~Londonderry, BT48 7JL, Northern Ireland, UK

    E-mail: k.wong-lin@ulster.ac.uk

**Acknowledgements**

This work was supported by the Alzheimer's Research UK (ARUK) Pump Priming Award (ARUK-2017NC-NI) (A.J., D.-H.W., S.W., P.L.M., T.S. and K.W.-L.), BBSRC (BB/P003427/1) (A.J., T.S., C.K.B. and K.W.-L), Ulster University Research Challenge Fund (C.K.B., P.L.M. and K.W.-L.), NSFC (31671077)(D.-H.W) and European Union's INTERREG VA Programme, managed by the Special EU Programmes Body (SEUPB (Centre for Personalised Medicine, IVA 5036)) (P. L.M. and K.W.-L.). K.W.-L. and S.W. were supported by COST Action (CA15120) Open Multiscale Systems Medicine

(OpenMultiMed) supported by COST (European Cooperation in Science and Technology). K.W.-L. was additionally supported by the Northern Ireland Functional Brain Mapping Project (1303/101154803) funded by Invest NI and Ulster University. The views and opinions expressed in this paper do not necessarily reflect those of the European Commission or the Special EU Programmes Body (SEUPB).

## Abstract

Alzheimer's disease (AD) is an age-specific neurodegenerative disease that compromises cognitive functioning and impacts the quality of life of an individual. Pathologically, AD is characterised by abnormal accumulation of beta-amyloid (A$\beta$) and hyperphosphorylated tau protein. Despite research advances over the last few decades, there is currently still no cure for AD. Although, medications are available to control some behavioural symptoms and slow the disease's progression, most prescribed medications are based on cholinesterase inhibitors. Over the last decade, there has been increased attention towards novel drugs, targeting alternative neurotransmitter pathways, particularly those targeting serotonergic (5-HT) system. In this review, we focused on 5-HT receptor (5-HTR) mediated signalling and drugs that target these receptors. These pathways regulate key proteins and kinases such as GSK-3 that are associated with abnormal levels of A$\beta$ and tau in AD. We then review computational studies related to 5-HT signalling pathways with the potential for providing deeper understanding of AD pathologies. In particular, we suggest that multiscale and multilevel modelling approaches could potentially provide new insights into AD mechanisms, and towards discovering novel 5-HTR based therapeutic targets.

## Introduction

Dementia is a clinical syndrome caused by a number of progressive illnesses that affect cognition, behaviour and the ability to perform daily activities (Weller and Budson, 2018). Dementia is one of the main causes of dependence and disability at older ages. Alzheimer's disease (AD) is the most common form of dementia affecting approximately 50 million people worldwide (Baldas et al., 2011). It has been estimated that by 2050, 115 million people worldwide will be living with AD (Wortmann, 2012).

Other types of dementia include vascular dementia, frontotemporal dementia, Lewy body dementia, Huntington's disease, and Creutzfeldt-Jakob disease, and co-morbidity of AD with some of these are not uncommon (Gale et al., 2018). An intermediate stage between healthy and AD is labelled mild cognitive impairment (MCI) (Gale et al., 2018). However, MCI is a loosely defined and heterogenous group, consisting of non-neurodegenerative or non-AD converters and people with other illnesses e.g. psychiatric illness (Gale et al., 2018).

To a large extent, AD can be categorized as familial AD (family history of the disease) or sporadic (late-onset) AD, with the latter overwhelmingly the most common type (Dorszewska et al., 2016). Various genes are currently thought to be associated with these different AD types. Mutations in amyloid precursor protein (APP), presenilin-1 (PSEN1) and presenilin-2 (PSEN2) are associated with familial AD while apolipoprotein E (ApoE) gene has been linked to the sporadic type (Gale et al., 2018). More generally, AD neuropathology has been characterised by the accumulation of β-amyloid (Aβ) protein due to the aberrant processing of APP, neurofibrillary tangles (composing of hyperphosphorylated tau protein), oxidative stress, excitotoxicity, neuroinflammation, and impairment in neurotransmitter systems (Butzlaff and Ponimaskin, 2016; Francis et al., 1999; Haruhiko et al., 2000; Heneka et al., 2015; Hynd et al., 2004; Markesbery, 1997; Rajmohan and Reddy, 2017).

Within the brain, acetylcholine (ACh) is an important neurotransmitter and neuromodulator implicated in cognitive functions such as learning and memory, and abnormalities (e.g., reduction in presynaptic ACh receptors, diminished choline acetyltransferase activity) in cholinergic neurons (which produce ACh) are found in the brains of AD patients (Francis et al., 1999). One approach to reducing the rate of cognitive decline in AD is to inhibit the breakdown of ACh into inactive metabolites by blocking the enzyme acetylcholinesterase responsible for the process. For instance, donepezil (Aricept) is a second-generation cholinesterase inhibitor (AChEI) and is the most widely prescribed drug for the treatment of AD. A Cochrane review estimated a 1.37 (95% CI: (1.13,1.61)) Mini Mental State Examination (MMSE) score improvement at six months after AChEI initiation (Birks and Harvey, 2018). However, it should be emphasised that these drug treatments do not cure AD, but rather delay the rate of cognitive decline associated with AD (Douchamps and Mathis, 2017). Hence, identifying novel therapeutics, especially repurposed drugs, has become a priority research area in both academia and industry.

AD has also been reported to be linked to changes in non-cholinergic neuromodulators, especially the monoaminergic systems (Morgese and Trabace, 2019). In particular, reduced serotonin (5-HT) levels in the neocortex and altered serotonin receptor (5-HTR) density are reported with severe cognitive decline in AD patients (Lai et al., 2005, 2002). Also, reduced 5-HT transporter (5-HTT) levels are reported in cortical, limbic, sensory, motor, striatal and thalamic brain regions in MCI as compared to healthy controls. Such reduction in 5-HTT levels is also found in the AD patient's temporal cortex (Smith et al., 2017; Tsang et al., 2003). These pathological conditions are possible grounds of underlying depressive symptoms, commonly seen in AD patients, and can potentially play an important role in the pathophysiology of AD (Smith et al., 2017).

Depression seems to be common in 20-30% of patients with AD (Tsuno and Homma, 2009). In fact, AD and depression share a close relationship: depression can lead to higher risk of AD but AD may also contribute to depression (Galts et al., 2019; Ownby et al., 2006). For instance, progression of AD is often associated with stress due to the catastrophic decline in motor and cognitive functions which can trigger the neural circuits involved in mediating stress response (Justice, 2018). Such chronic exposure of stress disrupts the cascades of stress hormones (e.g., cortisol) and affects the brain areas involved in monoaminergic transmission (e.g., dorsal raphe), decision making (e.g., prefrontal cortex), anxiety, and hence, increase the risk of developing depression (Arnsten, 2009; Bocchio et al., 2016; Hannibal and Bishop, 2014; Sengupta et al., 2017; Tafet et al., 2001).

Further, administration of antidepressants such as selective serotonin reuptake inhibitors (SSRIs) is often prolonged in chronically depressed patients, and long-term usage of SSRIs is associated with many side effects including cognitive impairments (e.g., memory deficits) (Prado et al., 2018; Sayyah et al., 2016). This gives rise to the possibility that extended use of SSRIs can increase the risk of developing dementia or AD. Practically, it is very challenging to establish a link between depressed individuals and late-life AD, as there are many constraints for long-term studies. Only a handful of studies attempted to establish such a link. For instance, a meta-study by Wang and colleagues suggests that antidepressant usage is substantially linked with a greater risk of developing dementia (Wang et al., 2018). These findings are not consistent with a study by Kessings and colleagues that suggests sustaining long-term treatment of SSRIs does not affect the risk of having dementia (Kessing et al., 2011). In contrast, a study by Xie and colleagues shows a pacified effect of SSRIs (e.g., fluoxetine) on cognitive performance in AD (Xie et al., 2019). Another study shows SSRIs to be beneficial in delaying the onset of AD on patients with a history of depression, a known risk factor for AD (Elsworthy and Aldred, 2019). Thus, these mixed results suggest that further studies are needed.

Amidst these mixed behavioural results, at the biological level, the effects are clearer. For example, the administration of SSRIs in AD in human studies (Lyketsos et al., 2003) and preclinical studies (animal models) have demonstrated a commendable influence of SSRIs on pathological markers of AD including A$\beta$ accumulation, tau deposits, and neurogenesis (Kim et al., 2013; Qiao et al., 2016; Sheline et al., 2014; Wang et al., 2014, 2016). As the administration of SSRIs, including in AD, can increase the level of 5-HT, which leads to the activation of several 5-HTRs (Tajeddinn et al., 2016; Tohgi et al., 1995), we shall next discuss in more detail the various effects of 5-HTR targeted drugs, with focus on 5-HT heteroreceptors, postsynaptic sites and AD.

## Serotonin receptor targeted drugs

5-HT receptors (5-HTR) are abundant throughout the brain with 14 known receptor subtypes and categorised into 7 subfamilies, 5-HT$_{1-7}$ (Stiedl et al., 2015). Apart from 5-HT$_3$R, all other 5-HT receptors act via G-protein coupled receptors (GPCR) (Hannon and Hoyer, 2002; Masson et al., 2012). 5-HT$_3$, the only 5-HT based ligand-gated ion channel, acts via changes in cation currents (e.g., Na$^+$, Ca$^{2+}$) (Masson et al., 2012). Numerous studies have associated 5-HT$_3$ mediated drugs with the accumulation of specific proteins found in AD (Fakhfouri et al., 2019; Huang et al., 2020; Skovgård et al., 2018). For example, in a rat model of AD, a study showed a protective response of a drug, tropisetron, a 5-HT$_3$ receptor antagonist, on A$\beta$-induced neurotoxicity on neurons in the hippocampus, a brain region associated with cognitive functions such as memory and spatial navigation (Rahimian et al., 2013). The hippocampus is known to be one of the regions of deterioration in early-stage AD (Gale et al., 2018). A follow-up study indicated that tropisetron protects rat pheochromocytoma cells (PC12) from oxidative-induced neurotoxicity via α7 nicotinic acetylcholine receptor (α7nAChR, which respond to ACh) and 5-HT$_3$R antagonist (Khalifeh et al., 2015). However, when a 5-HT$_3$ selective antagonist (ondansetron) is combined with acetylcholinesterase inhibitors (e.g., donepezil) in rats, they together potentiate and prolong the theta and gamma-band neural oscillations induced by donepezil alone. Such changes in hippocampal network oscillations and theta-gamma coupling are reported in a mouse model of AD, before the excessive production of Aβ (Goutagny et al., 2013; Skovgård et al., 2018).

In contrast to 5-HT$_3$R, the rest of the 5-HTR subtypes are associated with G protein-coupled receptors (GPCRs). In particular, the G-proteins for 5-HT$_1$R and 5-HT$_5$R act as members of the $G_{i/o}$ family of receptors (Masson et al., 2012) (Fig. 1). Activation of these receptors regulates various signalling pathways including protein kinase A (PKA) pathways (Fig. 1) (Masson et al., 2012). 5-HT$_1$R are expressed in large numbers in the hippocampus and are known to play a significant role in the regulation of memory processes (Ögren et al., 2008). Studies have also linked the level of 5-HT$_{1A}$R with agitative/aggressive behaviour (symptoms of AD), and the various stages of AD (Lai et al., 2003). Specifically, for the latter, a higher density of 5-HT$_{1A}$R was reported during the early stage of AD as compared to healthy controls (Truchot et al., 2008, 2007), whereas lower levels were detected during the advanced stage of AD (Becker et al., 2014; Kepe et al., 2006; Lanctôt et al., 2007; Vidal et al., 2016). In contrast, a post-mortem study indicated higher neocortical 5-HT$_{1A}$ density and relatively lower 5-HT levels in the brain tissue of AD patients as compared to healthy controls (Lai et al., 2002). There is also evidence that the drug Lecozotan, a 5-HT$_{1A}$R antagonist, can be used to improve cognitive dysfunction associated with AD, as the drug potentially interacts and enhance the signalling pathways of other neurotransmitters (e.g. cholinergic, glutamatergic) systems (Schechter *et al.*, 2005).

Unlike 5-HT$_1$R, 5-HT$_2$R are coupled to $G_q$ proteins and activate various signalling pathways including PKC and calcium/calmodulin-dependent kinase II (CaMKII), with the latter acting as a key protein kinase in neural plasticity and memory (Fig. 1).

Several studies have linked the density of 5-HT$_2$R with Aβ levels and cognitive impairment (Hasselbalch et al., 2008; Lai et al., 2005; Maroteaux et al., 2017; Nitsch et al., 1996; Štrac et al., 2016). In rats, intra-hippocampal injection of Aβ$_{1-42}$ reduces 5-HT$_{2A}$Rs and impairs memory (Holm et al., 2010). Not surprisingly, 5-HT$_{2A}$R expression levels were also reduced at the projection sites such as the medial prefrontal cortex (Christensen et al., 2008; Holm et al., 2010; Lorke et al., 2006). Additionally, the 5-HT$_2$R gene (T102C) is also reported to be involved in hallucination, psychosis and aberrant motor behaviour associated with AD (Tang et al., 2017). Recently, a study by Afshar and colleagues reported that 5-HT$_{2A}$R antagonist (NAD-299) individually, and in combination with 5-HT$_{1A}$R antagonist (TCB-2), significantly reduces oxidative stress and neuronal loss in hippocampal neurons in a rat model of AD (Afshar et al., 2019). Moreover, 5-HT seems to directly regulate Aβ via amyloid precursor proteins (APPs) via 5-HT$_{2A}$R and 5-HT$_{2C}$R with different signalling pathways as they lead to the formation of APPs and ultimately the accumulation of Aβ (Fig. 1); this effect is blocked by 5-HT antagonists (ketanserin, mianserin, and ritanserin) (Nitsch et al., 1996).

The other classes of 5-HTRs, 5-HT$_4$R, 5-HT$_6$R, and 5-HT$_7$R are coupled with G$_s$ proteins and activate various signalling pathways including protein kinase A (PKA) and extracellular signal-regulated kinase (ERK) (Fig. 1). Compared to other 5-HT receptors, 5-HT$_4$R targeted drugs have lately attracted considerable research interest, as many recent research studies have investigated its therapeutic potential in the treatment of AD (Bockaert et al., 2008). 5-HT$_4$R has been associated with learning and memory (Bockaert et al., 2008; Hagena and Manahan-Vaughan, 2017). 5-HT$_4$R expression has been found to be reduced in AD patients, while the activation of these receptors inhibits the (e.g., AC-cAMP-PKA) biochemical cascades that lead to AD (Fig. 1) (Hagena and Manahan-Vaughan, 2017; Reynolds et al., 1995).

Additionally, 5-HT$_4$R agonists have been found to improve cognitive deficits in AD (e.g., Claeysen, Bockaert and Giannoni, 2015). For instance, Madsen and colleagues have reported that cerebral 5-HT$_4$R binding is directly linked to abnormal accumulation of Aβ in AD patients (Madsen et al., 2011). Several other studies have also indicated that 5-HT$_4$R agonists (e.g., RS6733, a partial agonist) inhibit the production of Aβ in the entorhinal cortex, a region of deterioration in early-stage AD (Gale et al., 2018), by promoting the production of the neurotropic soluble APP alpha (sAPPα) and helps in improving cognitive abilities (learning and memory) in animal models of AD (Baranger et al., 2017; Yahiaoui et al., 2016). Further, RS6733 is also used with nicotinic receptor allosteric modulator/cholinesterase inhibitor galantamine to compensate for the deficit associated with short- and long-term memory (Freret et al., 2017). Interestingly, RS67333 also acts as an acetylcholinesterase (AChE) inhibitor and helps in reviving the cholinergic functions which are typically altered in AD (Lecoutey et al., 2014). Moreover, chronic administration of this drug decreases the levels of Aβ in the hippocampus in 5XFAD mouse model (expressing human APP and PSEN1 transgenes) of AD (Giannoni et al., 2013). However, when another 5-HT$_4$R agonist, SSP-002392, is applied to cultured human neuroblastoma cells, it increases sAPPα and cyclic adenosine monophosphate (cAMP) levels at a lower concentration than other well-known agonists (e.g., prucalopride), and suggesting the neuroprotective effect is mediated by EPAC (an exchange nucleotide protein directly activated by cAMP; Fig. 1) signalling (Cochet et al., 2013).

Apart from 5-HT$_4$R, 5-HT$_6$R has also attracted substantial research interest in the last few years. For instance, a 5-HT$_6$R antagonist, SB271046 has been found to recover memory impairment by reducing the levels of Aβ via inhibiting the gamma-secretase activity (multi-subunit enzyme that produces Aβ) in a mouse model of AD (Yun et al., 2015). Stimulation of these receptors primarily modulates the extracellular concentration of glutamate and GABA in various neural circuits and contributes to the release of other neuromodulators (e.g., dopamine, norepinephrine (Ne), ACh) which are known to be impaired in AD (Khoury et al., 2018). Similarly, another 5-HT$_6$R antagonist, idalopirdine (Lu AE58054), interacts with other neurotransmitter systems and increases the extracellular levels of dopamine, noradrenaline, and glutamate in the mPFC (de Jong and Mørk, 2017). Additionally, it has been suggested that these 5-HT$_6$R antagonists may have synergistic effects when combined with acetylcholinesterase inhibitors (e.g., donepezil), but recent phase III trials have reported no additional benefit or significant improvements in cognitive functions (Andrews et al., 2018).

As with the other 5-HTRs, 5-HT$_7$R is highly expressed in the hippocampus and plays an important role in memory formation, neuronal function and neurogenesis (Hashemi-Firouzi et al., 2017; Meneses, 2014; Shahidi et al., 2019). For example, activation of the 5-HT$_7$R via AS19, a selective serotonin agonist, improves synaptic impairment in a rat model of AD by decreasing apoptosis (programmed cell death) in the hippocampus (Hashemi-Firouzi et al., 2017) and could potentially hinder the progression of AD.

Additionally, glycogen synthase kinase (GSK-3) appears to be an important component in many 5-HT receptor-mediated signalling pathways (Fig. 1). Overactivity of GSK-3 is linked to familial and sporadic forms of AD in terms of increased levels of plaques and tangles (Proctor and Gray, 2010; Polter and Li, 2011). In general, GSK-3 is a ubiquitously present kinase that exists is two isoforms (GSK-3α and GSK-3β). Its activation largely depends upon the phosphorylation by upstream kinases including 5-HT mediated pathways (Fig. 1) (Lauretti et al., 2020; Polter and Li, 2011). It regulates key downstream biological pathways that are potentially involved in a range of diseases and disorders including cancer, diabetes, bipolar disorder, and neurodegeneration, and are often considered as a potential therapeutic target by many drug companies (Pandey and DeGrado, 2016; Saraswati et al., 2018).

For animal models of AD, several studies have reported that elevated GSK-3β activity is associated with increased levels of Aβ and tau hyperphosphorylation (Lauretti et al., 2020; Llorens-Martín et al., 2014). The functionality of GSK-3β can be regularised by phosphorylation/dephosphorylation that occurs at different sites (Lauretti et al., 2020). For example, phosphorylation on tyrosine-279/216 activates its activity while phosphorylation on serine 21/9 via different kinases (e.g., Akt, PKA) inhibits it (Sayas et al., 2006). Generally, GSK-3β activation is associated with generation and deposition of Aβ (Fig1). This is a multi-step process and involves modulation of the APP cleavage via different pathways (non-amyloidogenic, amyloidogenic) and includes synergic action of various enzymes. The non-amyloidogenic pathway of APP involves alpha-secretase (ADAM10, ADAM17) and gamma-secretase enzymes, and forms a degradable peptide (Lauretti et al., 2020). The amyloidogenic pathway includes the sequential action of beta-secretase (BACE-1) and gamma-secretase enzymes and constitutes intermediates: fibril and oligomers that eventually converts to Aβ (Lauretti et al., 2020; Llorens-Martín et al., 2014). In particular BACE-1 activity is elevated in AD patients (Decourt and Sabbagh, 2011). Further, studies have reported that inhibition of GSK-3β reduces BACE-1 mediated APP cleavage and ultimately reduces the Aβ levels (Lauretti et al., 2020; Ly et al., 2013). GSK-3 inhibitor (SAR502250) also provide neuroprotective effect in the animal model of AD (Griebel et al., 2019). Additionally, GSK-3β plays a key role in tau phosphorylation. GSK-3β phosphorylates at different sites including Thr231 and leads to the separation of microtubules that fosters the generation of tau oligomers and neurofibrillary tangles (NFTs) (Lauretti et al., 2020). Given the above results, GSK-3 inhibitors (e.g., Tideglusib) are currently undergoing phase II clinical trials (Griebel et al., 2019).

There is also evidence that 5-HT based drugs (e.g. SSRIs) interact with GSK pathways, and are effective in lowering some of the proteins that are impaired during AD. Specifically, the SSRI escitalopram was found to reduce $A\beta_{1-42}$ induced hyperphosphorylation of tau via the 5-HT$_{1A}$R mediated Akt/GSK-3β pathway in the hippocampal neurons (Wang et al., 2016) (Fig. 1) and this could be a key pathway for the potential treatment of AD, especially in its early stage. Furthermore, SSRI can also modulate GSK-3β signalling to regulate the neurogenesis in hippocampus neurons via activation of 5-HT$_{1A}$R (Fig.1 ) (Hui et al., 2015). Overall, these studies suggest that SSRIs and 5-HTR based drugs, and their combinations with other targeted neuromodulator (e.g. ACh) receptors, have the potential to perturb downward signalling pathways involve in the regulation of Aβ and tau. Currently, novel drug therapeutics for AD targeting the 5-HT system is the subject of intense research given that these drugs are already available and currently undergoing Phase II clinical trials (Šimić et al., 2017).

However, the system can be rather complex. For instance, it is known that any postsynaptic neuron can manifest a combination of two to three 5-HT receptor subtypes (Mengod et al., 2010). Thus, a pressing issue for the research community is to understand whether 5-HT induced signalling pathways crosstalk with other pathways which are impaired in AD. Given that 5-HT receptor subtypes can also respond with different affinities (Mengod et al., 2010), drug effects can lead to variable activation of intracellular signalling cascades. The complexity is compounded by the involvement of a multitude of proteins, enzymes, transporters, related genes, neuronal and synaptic properties, and ultimately cognition and behaviour – a multiscale and multilevel problem. Thus, the systems' complexity has hindered a deeper understanding of the underlying pathophysiological mechanisms that give rise to the abnormal accumulation of Aβ and tau and cognitive decline in AD has yet to be achieved.

Computational modelling offers a platform to bridge such a gap and guide drug design and development, and treatment (Roy, 2018). One way this can be achieved is through the integration of data or information across several experiments and the development of biologically based computational models. Indeed, such models have the potential to allow systematic exploration across multiple scales of description, beyond current experimental capabilities (Wong-Lin et al., 2017). This shall be our next point of discussion.

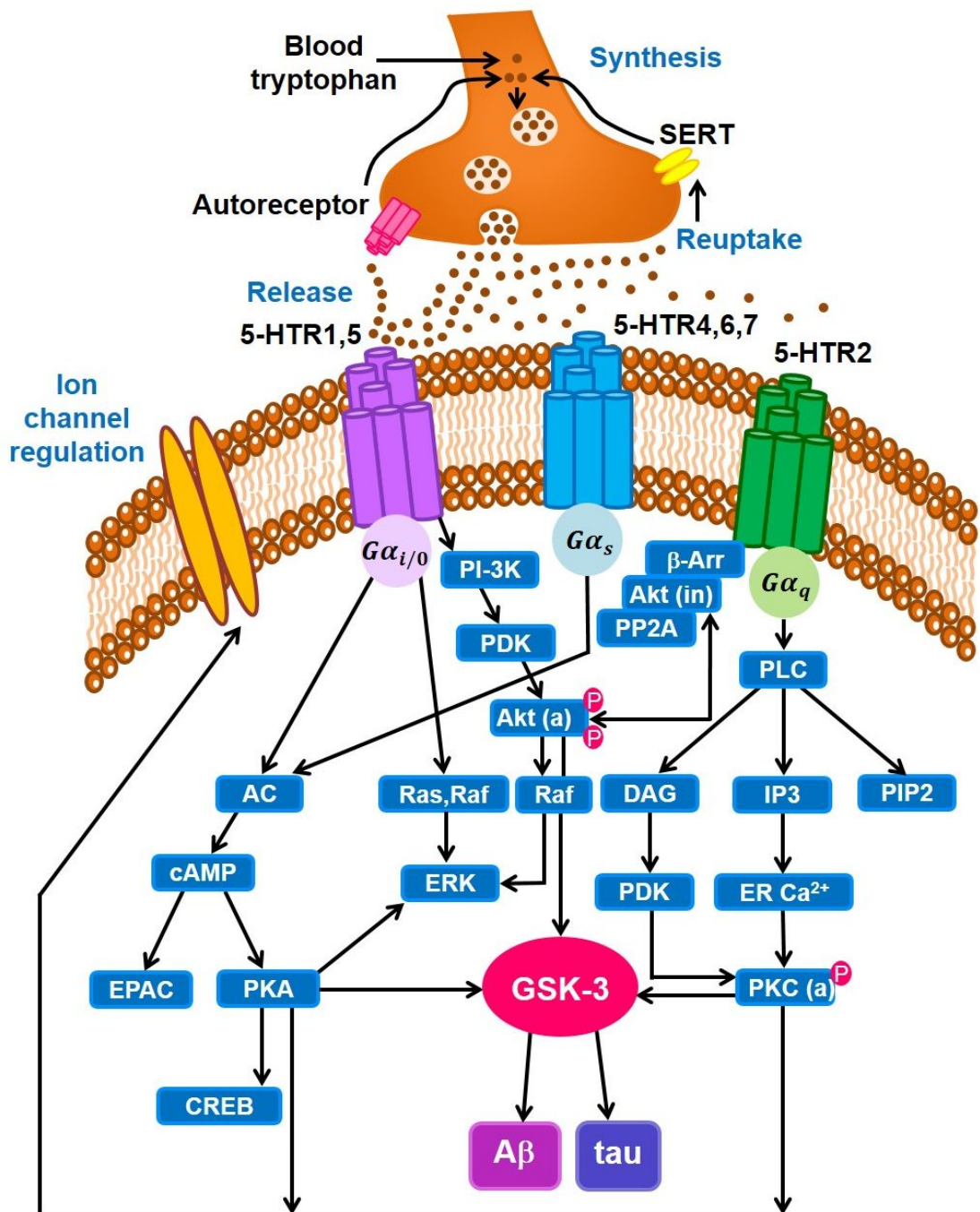

Figure 1. Schematic diagram of 5-HTR mediated signalling pathways: Activation of 5-HT$_{1A}$R initiates several pathways including G$\alpha_{i/o}$-adenylyl cyclase-cAMP-EPAC/PKA and Ras, Raf-ERK signalling pathways. When cAMP binds to the subunits of PKA, they phosphorylate downstream proteins to regulate key cellular processes. As a result, it increases the gene transcription involved in the development of long-term memory which is regulated by CREB. Furthermore, it alters the NMDA, AMPA and GABA receptor-mediated currents and also plays an active part in modulation of voltage-gated Na$^+$, K$^+$ and Ca$^{+2}$ ion channels. Additionally, 5-HT$_1$R activates Akt via PI-3K-PDK pathways. It is a key protein kinase and also regulates phosphor-Ser9-GSK3β. Similarly, activation of 5-HT$_2$R initiates PLC-DAG-PDK-PKC/IP3-ERCa$^{+2}$-PKC pathways. Importantly, the activated form of PKC also regulates phosphor-Ser9-GSK3β. Also, 5-HT$_{2R}$ affects β-Arrestin-Akt-PP2A signalling pathway. 5-HT: 5-hydroxytryptamine; SERT: serotonin reuptake transporter; G$\alpha_{i/o}$, G$\alpha_s$, G$\alpha_q$: isoforms of the α subunits of G protein-coupled receptors (GPCR); AC: adenylyl cyclase; cAMP: cyclic adenosine monophosphate; EPAC: exchange proteins activated by cAMP; PKA: protein kinase A; CREB: cAMP response element-binding protein; Raf: rapidly accelerated fibrosarcoma kinase; ERK: extracellular signal regulated kinase; PI-3K: phosphoinositide 3-kinases; PDK1: phosphoinositide-dependent kinase1; Akt (a): protein kinase B (active); β-Arr: β-arrestin; Akt (in): Protein Kinase B (inactive); PP2A: protein phosphatase 2; PLC: phospholipase C; DAG: diacylglycerol; IP3: inositol 1,4,5-trisphosphate; PIP2: phospholipid phosphatidylinositol 4,5-bisphosphate; ER: endoplasmic reticulum; PKC: protein kinase C; GSK-3: glycogen synthase kinase-3, there are two isoforms GSK-3α, GSK-3β; tau: tau protein; Aβ: beta amyloid; P: Phosphorylated. Note: Signalling pathway specific to 5-HT metabotropic receptors are shown; for detailed signalling networks, see (Masson *et al.* 2012) and (Wong-Lin *et al.* 2017), and for their association with GSK, see (Polter and Li 2011).

## Towards multiscale computational modelling of serotonergic system

Computational modelling or *in silico* investigation in preclinical and clinical research is an important research component towards facilitating understanding of brain functions and diseases such as AD, and for drug discovery and development (Cutsuridis and Moustafa, 2017a, 2016, 2017b; Geerts et al., 2017b; Roy, 2018). For years, mechanistic models have been developed to provide insights into the mechanism underlying the disease, to explore novel drug targets, and to gain a deeper understanding of drug actions (De Witte et al., 2016; Meier-Schellersheim et al., 2019; Schmidt et al., 2013). In particular, the key attractiveness of a quantitative model, captured by mathematical or statistical measures, is to provide a more integrated and quantitative understanding of mechanisms and patterns while eliminating ambiguities and adding rigour to "mental" models (for intuitive understanding) in experimental/clinical neurosciences. As in the field of physical sciences and engineering, experimental "what-if" scenarios can be tested in model simulations to evaluate hypotheses. Computational models also provide testable model predictions, which can guide future experiments (Mazein et al., 2018).

Several computational models of AD pathologies, symptoms and treatments have been proposed, and these are covered in recent comprehensive reviews (see e.g. Cutsuridis and Moustafa, 2017a, 2017b; Hassan et al., 2018, and references therein). Several earlier modelling papers in systems and theoretical biology were emphasised more towards understanding the dynamics of Aβ accumulation and its interactions with other proteins (Hassan *et al.* (2018)). Later modelling work encompassed the mapping of much larger number of biochemical interactions and other more data-driven, using either logic-based (Maude Petri net) models (Anastasio, 2011). For instance, such models could link Aβ effects on synaptic plasticity (Anastasio, 2014), and the search for potential combination of drugs to reduce microglial inflammation (Anastasio, 2015). More abstract probabilistic graphical (Bayesian) network models were also applied to understand key protein/drug interactions at the systems level (Rembach et al., 2015).

To date, there are only a small number of computational modelling studies on 5-HT mediated signalling pathways. Importantly, none of the computational studies are focused on understanding the role of 5-HT mediated intracellular signalling pathways in AD. Hence, these present opportunities for computational modellers to contribute. Thus, we shall henceforth first discuss a set of models on 5-HT mediated pathways, before discussing a separate set of models on aggregation of Aβ and hyperphosphorylated tau and tangles. Then, we argue the potential to integrate these two sets of models together.

In a highly detailed computational model of 5-HT receptor-mediated signalling, Chang *et al.* (2009) developed a model of 5-HT$_{1A}$R and 5-HT$_{2A}$R activated ERK(1/2) pathways using Michaelis–Menten formalism and the law of mass action. In the model, 5-HT$_{1A}$R stimulated phosphoinositide 3-kinases (PI-3K) pathway while 5-HT$_{2A}$R triggered mitogen-activated protein kinase (MAPK) / ERK pathway (also known as the Ras-Raf-MEK-ERK pathway) (Fig. 1). Their model's key results, in agreement with experimental data, showed the dominance of 5-HT$_{2A}$R over 5-HT$_{1A}$R in the MAPK signalling pathway, and the deleterious effects of regulator/enzymes affecting basal levels of ERK. In another modelling work, Zhang *et al.* (2012), building on the model by Pettigrew et al. (2005) (Pettigrew et al., 2005; Zhang et al., 2012), studied the effects of 5-HT on PKA-ERK interactions to enhance long-term facilitation of synapses. In a similar vein, Zhou et al. (2014) modelled both PKA and PKC signalling to show that PKC was sufficient for short-term facilitation of synapses, and that cooperation among the signalling cascades could potentially contribute to the enhancement of learning and memory, which had recently been validated experimentally (Liu et al., 2017; Zhou et al., 2014). These results may have implications in AD, given the latter's deterioration in learning and memory. Importantly, these models also laid the foundation for modelling 5-HT signalling pathways through PKA and PKC responses (Fig. 1).

As discussed earlier, and illustrated in Fig. 1, 5-HT signalling not only involves ERK and PKA pathways but also engages the GSK-3 enzymes via 5-HTRs. Also, GSK-3 can in turn modulate 5-HT$_{1B}$R, which exists at presynaptic terminals of 5-HT neurons, leading to changes in 5-HT level. GSK-3 is inactivated by serine phospho-ser21/Ser9 (ps21/9), and the latter bridges between 5-HTRs and GSK-3 (Fig. 1). The intermediate transcription factor p53, which regulates the expression of cellular stress response genes, can interact with GSK-3 (Jazvinšćak Jembrek et al., 2018). Mild oxidative stress injury can lead to p53 ensuring antioxidative activities and promoting cell survival. But over antioxidative capacity can lead to cell death by p53 (apoptosis). Hence, this results in a possible link between 5-HT, GSK-3 and the fields of neuroinflammation and immunology.

Such biochemical reaction process involving GSK-3, p53, tau tangles and Aβ aggregation was first modelled by Proctor and Gray (2010) using Systems Biology Markup Language (SBML) and stochastic simulation, building on their previous model of p53 (Proctor and Gray, 2008). Their modelling results accounted for the overactivity of GSK-3 and p53 after a stress event, leading to increase in Aβ and tau tangles, providing a correlation between the latter two, but not causation. Later, Proctor et al. (2013) extended the model and showed that immunisation helped to clear plaques, but limited influence on soluble Aβ, phosphorylated tau and tangles, consistent with experimental observation (Proctor et al., 2013). Interestingly, the model results suggested interventions to be performed at a very early stage of AD. It should also be noted that GSK-3 is also associated with other neurodegeneration related to tauopathies, such as Pick's diseases, progressive supranuclear palsy and corticobasal degeneration (Ferrer et al., 2002). Hence, it may be worth investigating through computational modelling to further understand the 5-HT based drug (e.g., SSRI) effects and treatments on these other diseases.

Taken together, despite their small number, the abovementioned mechanistic models were used as examples which could potentially be integrated to provide a deeper understanding of the effects of 5-HT through GSK-3 to aggregation of plaques and tau tangles and subsequent changes in neuronal and synaptic properties, and potential novel therapeutics. For instance, 5-HT induced PKA and PKC pathways can modulate at least four membrane currents including 5-HT sensitive, voltage-dependent and calcium-activated potassium currents and L-type calcium currents (Baxter et al., 1999). Notably, modulation of voltage-dependent potassium currents by PKC plays a significant part in the broadening of spikes (Baxter et al., 1999). In contrast, simultaneous modulation of 5-HT sensitive potassium and calcium current increases the excitability within cells, potentially due to the indirect crosstalk among PKC and PKA pathways. Further, Aβ can interact with membrane ion channel currents and can affect their overall excitability. For example, computational studies by Zou and colleagues showed that Aβ blocked A-type currents and increased the excitability of pyramidal neurons, and subsequently the excitability of the (hippocampal septal) microcircuit (Zou et al., 2012, 2011). This could in turn lead to memory impairment and even symptoms of epileptic seizures, a common comorbidity in AD (Zou *et al.*, 2012). Complementary to these studies, Abuhassan and colleagues used a large-scale model to understand the effects of synaptic loss, due to AD, on global oscillatory dynamics (Abuhassan et al., 2014). Thus, there is the potential for computational models to bridge from one level of description to another – multiscale and multilevel modelling. However, linking from one modelling scale/level to another requires its abstraction (Wong-Lin et al., 2017).

In fact, a challenge to mechanistic models lies in how they can be utilised when certain components of the network are not parameterised or missing, which can impact on model accuracy (Fröhlich et al., 2018). A practical way to estimate the optimal set of model parameter values is to fix certain known model parameters and manipulate others within a physiologically reasonable range until the model is able to mimic experimentally observed pattern(s) (Maex et al., 2009; Sterratt et al., 2011). However, as the number of unknown parameters increases, typically in the context of a larger signalling network, it becomes exponentially more difficult to systematically search the parameter space and unique solutions are not guaranteed, since there may exist many combinations of parameters that can furnish similar outputs. An alternative approach to dealing with such challenges is creating simpler, reduced computational or mathematical models that approximate the larger network (Albert and Thakar, 2014).

One approach is to reduce the complexity of the dynamic behaviour of the model that can be introduced when the model exhibits bi- or multi-modality is to use Boolean models. Such models can qualitatively recreate the temporal dynamics of a larger signalling network (Albert and Thakar, 2014). These models are particularly useful when limited kinetic details about the interaction of components are available (Albert and Thakar, 2014). For instance, when Boolean models are used in a gene regulatory network, they can describe the characteristics of circadian systems (Akman et al., 2012; Watterson and Ghazal, 2010). These models are also employed to analyse the influence of stress and SSRIs in complex networks of 5-HT, neurotrophin and cortisol mediated signalling pathways (Moreno-Ramos et al., 2013). Notably, one such model predicts the network dynamics, especially when specific genes are knocked out (Moreno-Ramos et al., 2013).

Another promising approach to minimise the number of unknown parameters in mechanistic models is to estimate the prior knowledge of the unknown components. This can be achieved by using Bayesian methods (Spiegelhalter et al., 2002). For example, Bayesian methods are used to deduce mechanistic parameters for amyloid formation kinetics (Nakatani-Webster and Nath, 2017). Other methods to reduce the parameter space in signalling pathways include perturbation techniques. For instance, Flower and Wong-Lin (2014) and Cullen and Wong-Lin (2015) used step perturbation technique to elucidate key model components (e.g. substrates) and their temporal dynamics, which lead to substantial reduction of model sizes for intracellular signalling in presynaptic terminals of 5-HT- and dopamine-producing neurons, respectively. Perhaps similar or more advanced techniques could be applied to models of postsynaptic 5-HT mediated signalling pathways.

## Conclusion and future directions

AD is a complex neurodegenerative disorder characterised by cognitive impairment comorbid with behavioural changes that considerably affect day-to-day functioning. Neuropathologically, AD is marked by an excessive accumulation of Aβ and hyperphosphorylated tau protein. Currently available drugs for AD are primarily used to reduce symptoms or control behaviour, but not cure AD. Majority of them target neurotransmitter systems that include cholinergic, non-cholinergic, glutamatergic and their combinations.

The focus of our review is to highlight the role of 5-HT system in AD. We discussed the 5-HTR mediated signalling pathways. These pathways are targeted by drugs such as SSRIs. We then discussed the role of SSRIs and 5-HTR mediated drugs in AD. Newer generation of antidepressant drugs such as SNRIs (e.g., venlafaxine) can provide an alternative route for the treatment of AD, as there is emerging evidence that suggests that norepinephrine system is also involved in Aβ regulation (Liu et al., 2015; Mokhber et al., 2014; Ross et al., 2015). Hence, further modelling work on the interaction of neuromodulators will potentially be enlightening (Jalewa et al., 2014; Joshi et al., 2017, 2015, 2011; Wang and Wong-Lin, 2013).

We then highlighted the importance of the GSK-3 protein kinase, as a key kinase that sits in the downstream signalling pathway of most of the 5-HT receptors especially 5-$HT_{1A}R$ and $5HT_{2A}R$. We then discussed existing computational models that describe the mechanisms that link GSK-3 to aggregation of Aβ and tau hyperphosphorylation. Then, as an example, we suggested that both types of models could potentially be integrated to understand novel 5-HT based therapeutics for AD. Further, we highlighted the importance of reduced modelling approaches such as Boolean and Bayesian approaches, especially when there are many unknown parameters in a model or when the level of description in the data varies a lot. Abstraction of these models could be used as basis for models in adjacent scale, e.g. neuronal or neuronal network model.

Taken together, we have shown that there are currently not many multiscale computational models of 5-HT mediated signalling pathways and their links to AD. Hence, there are ample opportunities for computational scientists or mathematical modellers in this research area. Collating and unifying models will become essential (Lloret-Villas et al., 2017).

As hinted earlier, complementing mechanistic modelling approaches, we can also utilise knowledge- and data-driven approaches (Geerts et al., 2017a; Li et al., 2009). Compared to mechanistic models, non-mechanistic, data-driven computational models can forge relationships among the input and output datasets, without taking into account the underpinning biological processes (Zhang et al., 2018). These methods largely use probabilistic or statistical methods, including machine learning approaches, to solve complex problems, and a key advantage is that they could work well with highly heterogeneous datasets (Ding et al., 2018). Knowledge-driven approaches make use of available literature, clinical/medical records, and online resources to mine relevant information (Younesi and Hofmann-Apitius, 2013). Usually, these models operate in conjunction with data-driven models and can feed to other types of models (e.g., mechanistic), for example, to identify correlation or (probabilistic) causality for a better understanding of the relationship among the system's components (e.g., proteins, enzymes) present in the intracellular signalling network (Younesi and Hofmann-Apitius, 2013).

These types of models are highly relevant and can be used in conjunction with mechanistic models to understand disease mechanism(s) and potential novel therapeutic approaches. For instance, with such models, we could explore how drug(s) can activate 5-HTR mediated signalling pathways which involve GSK-3 protein kinase and regulates A$\beta$ and tau. Not surprisingly, interaction and parameter values of these pathways may be different for different brain regions. Practically, it is impossible to estimate the activity of these signalling pathways for all brain regions. Thus, there remains considerable opportunity for computational studies to estimate the activities of key proteins for different brain areas and link outcomes with properties that can be mechanistically modelled e.g. synaptic currents. These modelling features can then be a key ingredient to study how changes in signalling pathways can affect the function of neural networks via current modulation which in turn could affect cognitive (dys)function (Cano-Colino et al., 2013; Eckhoff et al., 2009).To achieve this, it is of utmost importance that experimentalists and computational modellers work together more synergistically.